\documentclass{aip-cp}

\usepackage[numbers,sort&compress]{natbib}
\usepackage{rotating}
\usepackage{graphicx}

\begin{document}

\title{Collective Phenomena in $pp$ and $ep$ Scattering}

\author[aff1,aff2]{Francesco Giovanni Celiberto}
\eaddress{francescogiovanni.celiberto@fis.unical.it}
\author[aff1]{Roberto Fiore}
\eaddress{roberto.fiore@fis.unical.it}
\author[aff3]{L\'aszl\'o Jenkovszky\corref{cor1}}

\affil[aff1]{Dipartimento di Fisica, Universit{\`a} della Calabria, 
              Arcavacata di Rende, 87036 Cosenza, Italy}
\affil[aff2]{Istituto Nazionale di Fisica Nucleare, 
              Gruppo Collegato di Cosenza, 
              Arcavacata di Rende, 87036 Cosenza, Italy}
\affil[aff3]{Bogolyubov Institute for Theoretical Physics, 
              National Academy of Sciences of Ukraine, Kiev, 
              03680 Ukraine}


\corresp[cor1]{Corresponding author: jenk@bitp.kiev.ua}

\maketitle

\begin{abstract}
Bjorken scaling violation in deep inelastic electron-proton scattering (DIS) is related to the rise of hadronic cross sections by using the additive quark model. Of special interest is the connection between saturation in the low-$x$ behavior of the DIS structure functions (SF) and possible slow-down of the $pp$ cross section rise due to saturation effects. We also identify saturation effects in the DIS SF with phase transition that can be described by the Van der Waals equation of state.     
\end{abstract}

\section{INTRODUCTION}
\label{introduction}
Although deep inelastic lepton-hadron scattering (DIS) 
is thought mainly as a ``microscope'' identifying 
individual point-like constituents in the hadron, 
collective effects may come into play with decreasing $x$ 
(Bjorken variable) and increasing photon virtuality $Q^2.$ 
In this paper we present two examples of such phenomena. 
The first example, given in the second Section, 
is a connection between expected saturation 
in the low-$x$, high-$Q^2$ DIS structure function (SF) 
and the high-energy behavior of the hadronic total cross section. 
The other one, given in the third Section, 
is a statistical physics interpretation of the DIS SF.

\section{BJORKEN SCALING VIOLATION AND RISING CROSS SECTIONS}
\label{sec:satur}
The main idea of this section is that the rise of hadron-hadron cross sections at high energies is related to the amount of Bjorken scaling violation at low $x$. 

By definition, the virtual photon-proton total cross section is related to the relevant DIS SF by
\begin{equation}\label{Eq:VdW0}
F_2(x,Q^2)=\frac{Q^2(1-x)}{4\pi\alpha(1+4m^2x^2Q^2)}\sigma_t^{\gamma^*p}\;,
\end{equation} 
$s=W^2=Q^2(1-x)/x+m^2.$

Real photon-proton total cross section is known~\cite{PDG} to be about two orders of magnitude below that of proton-proton cross section. 

In~\cite{JS} the $Q^2-$dependent structure function was related to on-mass-shall hadronic cross sections by means of dispersion relations and duality in $Q^2$,
\begin{equation}
  F_1(s,Q^2)=\int^{\infty}_{Q_0^2}\frac{\Phi(s,Q'^2)}{(Q'^2+Q^2)^2}dQ'^2\;,
  \;\;\;\;\; 2xF_1=F_2,
\end{equation} 
where the asymptotic part of $\Phi$ was saturated by $\rho$ and $\omega$ vector meson masses. 
With a simple phenomenological model of the SF, {\it e.g.}
\begin{equation}
F_2^{As.}(x,Q^2)=0.2\Bigl(1-0.05\ln\frac{Q^2}{Q^2_0}\ln\frac{x}{x_0}\Bigr)\;,
\end{equation} 
after integration in $Q^2,$ with lower integration limits corresponding to vector meson masses, 
and the relation $\sigma_{\rho N}\approx\sigma_{\omega N}\approx\sigma{\pi N}$, one gets a reasonable expression for the rising meson-baryon total cross section
\begin{equation}
  \sigma_{\pi N}(s)=24 \ \Bigl(1+0.1\ln\frac{s}{Q^2_0}\Bigr)\ {\rm mb}\;,
  \;\;\;\;\; Q^2_0=3 {\rm \ GeV}^2\;.
\end{equation} 

In the additive quark model, the hadron-hadron total cross section can be written as a product of the cross sections of the constituents, $\sigma_{qq}$~\cite{JS}, {\it e.g.}
\begin{equation}
\sigma(s)^t_{pp}=\sigma_{qq}[n_V+n_S(s)]^2\;,
\end{equation}
where $n_V$ is the number of valence quarks and $n_S(s)$ is that of sea quarks, their number increasing with energy.
 
We suggest that the increasing number of sea quarks is related to the
Bjorken scaling-violating contribution to the deep inelastic lepton-hadron structure function (DIS SF), namely to the momentum fraction of the relevant quarks given by the integral over the DIS structure function $F_2(x,Q^2)$. 

The number of quarks in a reaction can be calculated from the SF by means of sum rules. The momentum fraction carried by a parton is $\int_0^1dxF_2(x,Q^2),$
and hence the number of partons in the reaction is its inverse. Focusing on the global and asymptotic behavior of both the structure functions (at low $x$) and of the total cross sections (large $s$) and possible signs of the interrelated saturation effects, here we disregard details 
related to the flavor content of the hadrons.

A simple model for the DIS structure function results in the following expression for the total cross section, compatible with the data
\begin{equation}
\sigma(s)^t_{pp}=\sigma_{qq}n_V^2[1+0.016 \, \ln (s/Q^2_0)]\;,
\end{equation}  
where $\sigma_{qq}$ is a free parameter; $Q^2_0$ was fitted to the DIS data, and $n_V=3$.

Of special interest is the expected correlation between the onset of saturation in DIS (see, {\it e.g.}~\cite{Quarks, Lowx, Myronenko} and the possible slow-down of the rise of hadronic cross sections to follow). 

\section{DIS AS A VAN DER WAALS SYSTEM OF VIRTUAL PARTONS}
\label{sec:VdW}
\begin{figure}[h] 
 \label{Fig:VdW}
 \begin{minipage}{.5\textwidth}
  \centering
  \hspace{-1.0cm}
  \includegraphics[scale=0.425]{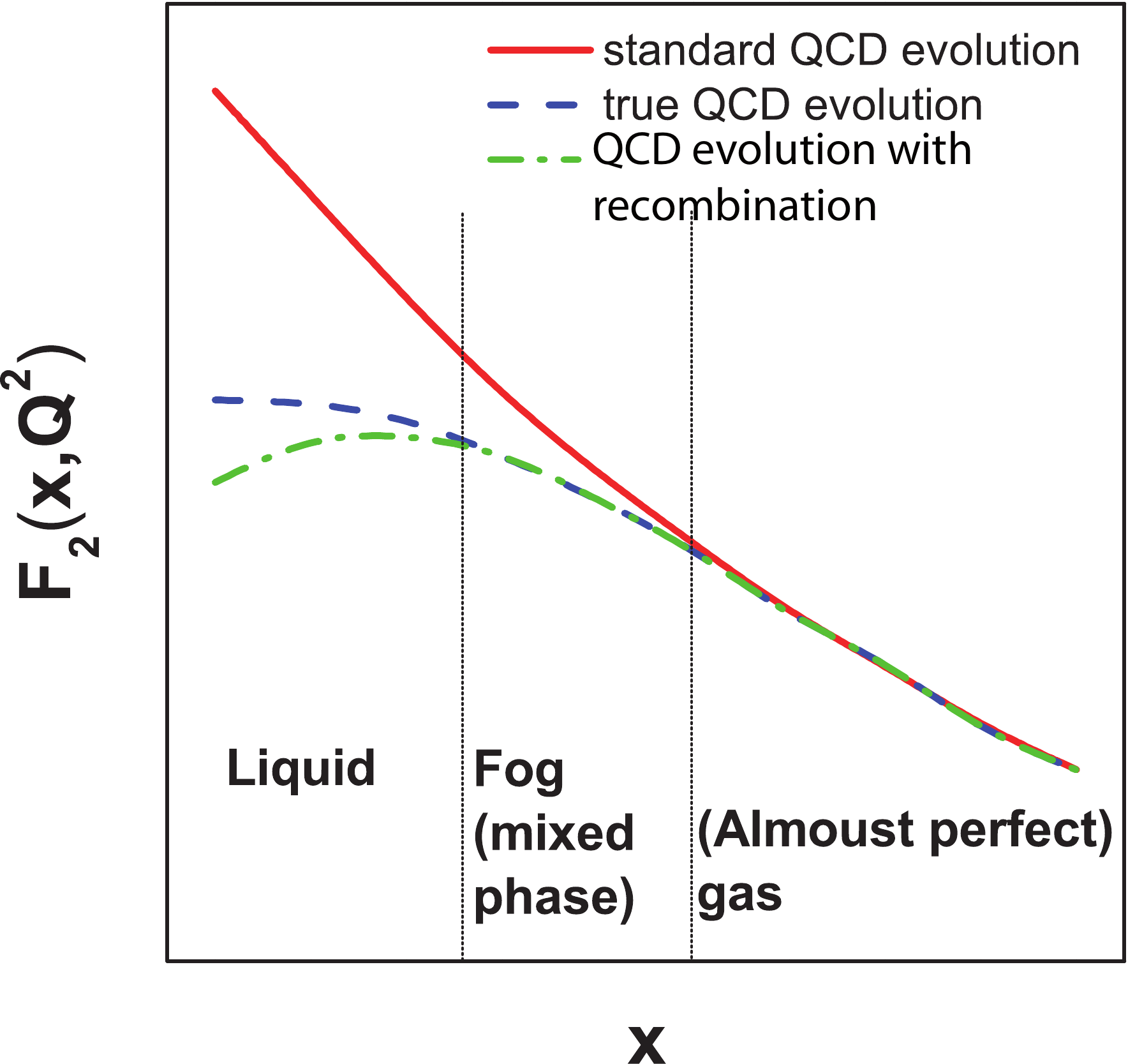}
 \end{minipage}
 \begin{minipage}{.5\textwidth}
  \centering
  \includegraphics[scale=0.54]{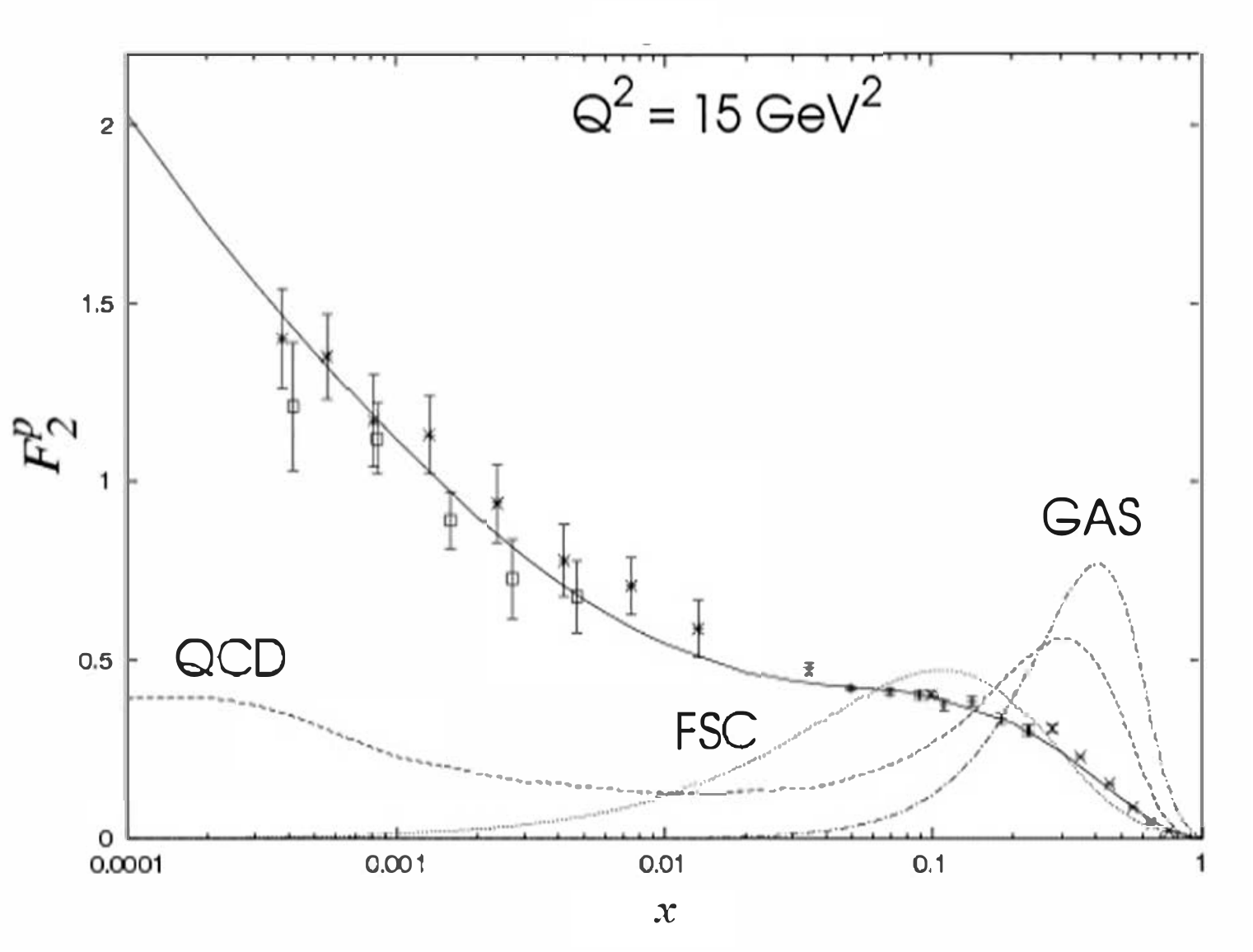}
  \caption{Left icon: Gaseous, foggy and liquid states in various kinematic regions of DIS (schematic). Right icon: Same as on the left, but with the data at $Q^2=15$ GeV$^2$ also shown.}
 \end{minipage}
\end{figure}

It was suggested in~\cite{VdW1,VdW2} that saturation in DIS, predicted by QCD and observed experimentally, corresponds to the condensation of the partonic gas to a fluid. The interior of the nucleon exited in DIS or DVCS undergoes a phase transition from a parton gas (high- and intermediate $x,\  \  x\sim 0.05$) to a partonic fluid. The division line is located roughly at those values of $x$ and $Q^2$ where Bjorken scaling is violated, as shown in Figure~\ref{Fig:VdW}, left icon.

Formally, saturation may be treated in the context of the BFKL equation~\cite{BFKL}
\begin{equation}
\frac{\partial N(x,k_T^2)}{\partial\ln(1/x)}=\alpha_sK_{BFKL}\otimes N(x,k_T^2)\;,
\end{equation}
where $K_{BFKL}$ is the BFKL integral kernel (splitting function). Since the number of partons increases with energy, at certain ``saturation" value of $x$ and $Q^2$, an inverse process, namely  recombination of pairs of partons comes into play, and the BFKL equation is replaced by  
\begin{equation}
\frac{\partial N(x,k_T^2)}{\partial\ln(1/x)}=\alpha_sK_{BFKL}\otimes N(x,k_T^2)-\alpha_s[K_{BFKL}\otimes N(x,k_T^2)]^2\;,
\end{equation}
based on the simple idea that the number of recombinations is roughly proportional to the number of
quark parton pairs, $N^2$. Saturation sets in when the second, quadratic term overshoots the first, linear one, thus tempering the increase of the number of produced partons and securing unitarity. 

The transition between the partonic gas, via a mixed foggy phase, to the partonic liquid (see Figure~\ref{Fig:VdW}) may quantified by the Van der Waals equation of state (EoS)
\begin{equation}\label{Eq:VdW}
(P+N^2a/V^2)(V-Nb)=NT\;,
\end{equation}  
where $a$ and $b$ are parameters depending on the system, $N$ is the number of particles and $V$ is the volume of the ``container", $V(s)=\pi R^3(s),\ \ R(s)\sim\ln s$ being the nucleon radius. For point-like particles (perfect partonic gas) $a=b=0$ and the VdW equation reduced to $pV=NT$, and, since $N/V\sim T^3,$ we get in this approximation $p\sim T^4.$ 

Alternatively, the Van der Waals equation may be written as
\begin{equation}\label{Eq:VdW1}
(P+a/V^2)(V-b)=RT\;,
\end{equation} 
or, equivalently, 
\begin{equation}\label{Eq:VdW2}
P=\frac{RT}{V-b}-a/V^2\;.
\end{equation} 
The parameter $b$, responsible for the finite dimension of the constituent, in our case is $\sim 1/Q,$ while the term $a/V^2$ is related to the long-range forces between the constituents. From this equation one finds for critical values, $V=V_c,\ \ P=P_c,\ \ T=T_c$ in terms of the parameters $a$ and $b$:  
\begin{equation}
V_c=3b,\ \ P_c=a/(27b^2),\ \ T_c=8a/(27Rb)\;.
\end{equation} 

The number of particles may be calculated as
\begin{equation}
N(s)=\int_0^1 dx F_2(x,Q^2)\;,
\end{equation} 
where $F_2(x,Q^2)$ is the nucleon SF, measured in DIS.

\section{CONCLUSIONS}
\label{conclusions}
Collective properties of the nuclear matter are studied mostly in heavy-ion collisions. In this paper we presented examples of collective effects in DIS. As energy is pumped into a nucleon by a virtual photon, the number of partons populating it increases, reaching critical densities, thus causing phase transitions. The difference between the case of hadronic (heavy nuclei) collisions and DIS is that in the latter case the nucleon (nucleus) is heated by a highly virtual photon (heavy vector meson), while in heavy-ion (nucleon) collisions we deal with on-shall (real) objects. However, the produced hot and dense matter is of the same nature.    

\section{ACKNOWLEDGMENTS}
\label{acknoledgments}
We thank the organizers, for their hospitality and for the inspiring atmosphere at the conference. L.J. acknowledges the financial support during the conference.

\end{document}